\documentclass[twocolumn,showkeys,showpacs,preprintnumbers,prd,superscriptaddress,nofootinbib]{revtex4-1}
\usepackage{graphicx}
\usepackage{epsf}
\usepackage{bm}
\usepackage{amsmath}
\usepackage{amsfonts}
\usepackage{amssymb}
\usepackage{epstopdf}
\usepackage{natbib}
\usepackage{hyperref}
\usepackage{color}
\usepackage{verbatim}
\usepackage{multirow}
\usepackage{bm}
\usepackage{hyperref}
\usepackage{float}

\definecolor{darkblue}{rgb}{0.0, 0.0, 0.55}
\definecolor{darkred}{rgb}{0.55, 0.0, 0.0}

\usepackage{hyperref}
\hypersetup{
    colorlinks=true, 
    linkcolor=darkblue,
    citecolor=darkblue,
    urlcolor=darkblue}
    
\makeatletter\let\expandableinput\@@input\makeatother

\begin{document}

\title{Sign Switching in Dark Sector Coupling Interactions as a Candidate for Resolving Cosmological Tensions}


\author{Miguel A. Sabogal}
\email{miguel.sabogal@ufrgs.br}
\affiliation{Instituto de F\'{i}sica, Universidade Federal do Rio Grande do Sul, 91501-970 Porto Alegre RS, Brazil}

\author{Emanuelly Silva}
\email{emanuelly.santos@ufrgs.br}
\affiliation{Instituto de F\'{i}sica, Universidade Federal do Rio Grande do Sul, 91501-970 Porto Alegre RS, Brazil}

\author{Rafael C. Nunes}
\email{rafadcnunes@gmail.com}
\affiliation{Instituto de F\'{i}sica, Universidade Federal do Rio Grande do Sul, 91501-970 Porto Alegre RS, Brazil}
\affiliation{Divisão de Astrofísica, Instituto Nacional de Pesquisas Espaciais, Avenida dos Astronautas 1758, São José dos Campos, 12227-010, São Paulo, Brazil}

\author{Suresh Kumar}
\email{suresh.kumar@plaksha.edu.in}
\affiliation{Data Science Institute, Plaksha University, Mohali, Punjab-140306, India}

\author{Eleonora Di Valentino}
\email{e.divalentino@sheffield.ac.uk}
\affiliation{School of Mathematical and Physical Sciences, University of Sheffield, Hounsfield Road, Sheffield S3 7RH, United Kingdom} 


\begin{abstract}
The $\Lambda$CDM model has successfully explained a wide range of cosmological observations, but is increasingly challenged by the emergence of cosmological tensions, particularly the Hubble Tension $H_0$ and the $S_8$ tension. The Hubble Tension, with a significance above 5$\sigma$, and the $S_8$ tension, showing a discrepancy of approximately 2-4$\sigma$, highlight inconsistencies between measurements of the local and early universe. 
This paper expands a well-established Interacting Dark Energy (IDE) phenomenological scenario, where dark matter (DM) can transfer energy to dark energy (DE) or vice versa, depending on the sign of the coupling parameter $\xi$. The novel feature consists in a transition mechanism which reverses the direction of the energy-momentum transfer after the redshift where the densities of the dark species are the same. We evaluate this model using a comprehensive set of recent observational data, including Baryon Acoustic Oscillations (BAO) from the DESI survey, Type Ia Supernovae from the PantheonPlus, DESY5 and Union3 samples, and Cosmic Microwave Background (CMB) data from Planck. 
Our analysis shows that this scenario can potentially relax both the $H_0$ and $S_8$ tensions simultaneously. We find the new model to be weakly preferred over $\Lambda$CDM by BAO-DESI data. However, we show that the IDE model features positive Bayesian evidence compared to $\Lambda$CDM only when Cepheid distance calibration in the SH0ES sample is used to calibrate SNIa data from PantheonPlus. 

\end{abstract}



\maketitle

\section{Introduction}

The standard cosmological model ($\Lambda$CDM) has performed well in explaining observations made over recent decades~\cite{Planck:2018nkj,Planck:2018vyg,Planck:2019nip,Mossa:2020gjc,ACT:2020gnv,eBOSS:2020yzd}. However, with the improvement in measurement precision, several gaps have emerged that this model cannot address. The most statistically significant of these concerns the rate of cosmic expansion today. Probes of the local universe provide higher values for the Hubble constant $H_0$, while probes of the early universe assuming a $\Lambda$CDM scenario indicate lower values~\cite{Riess:2021jrx,Verde:2019ivm,Knox:2019rjx,DiValentino:2020zio,DiValentino:2021izs,DiValentino:2022fjm,Kamionkowski:2022pkx,Verde:2023lmm,DiValentino:2024yew,Breuval:2024lsv,Li:2024yoe,Murakami:2023xuy,Said:2024pwm,Boubel:2024cqw,Freedman:2024eph,Riess:2024vfa,Scolnic:2024hbh}. This discrepancy, known as the Hubble Tension, currently has a significance above 5$\sigma$~\cite{Riess:2021jrx,Breuval:2024lsv}. Additionally, there is a growing tension in the parameter $S_8$, which relates the amplitude of matter fluctuations to the total matter density parameter ($\Omega_{\rm m}$) via $S_8 = \sigma_8 \sqrt{\Omega_{\rm m}/0.3}$. Measurements from the early universe, such as those from Planck CMB, indicate a higher value of $S_8$ compared to values obtained from weak gravitational lensing, galaxy cluster, and cluster counts surveys, with a discrepancy of approximately 2-4$\sigma$~\cite{DES:2021wwk,DiValentino:2020vvd,DiValentino:2018gcu,Kilo-DegreeSurvey:2023gfr,Troster:2019ean,Heymans:2020gsg,Dalal:2023olq,Chen:2024vvk,ACT:2024okh,DES:2024oud,Harnois-Deraps:2024ucb,Dvornik:2022xap,Armijo:2024ujo}. In addition to the aforementioned issues, recent literature highlights other anomalies and discrepancies in the $\Lambda$CDM framework~\cite{Rogers:2023upm, Nunes:2021ipq, Ruchika:2024ymt,Lopez-Corredoira:2024pgl,Pourojaghi:2024bxa,Risaliti:2023uiy,Green:2024xbb,Giare:2023wzl,Jiang:2024viw,DiValentino:2021izs,Abdalla:2022yfr,Perivolaropoulos:2021jda,Khalife:2023qbu,DiValentino:2024yew,Givans:2023kbg,Hazra:2024nav,Craig:2024tky,Carniani_2024,DESI:2024mwx,Giare:2022rvg,RoyChoudhury:2024wri}.

Motivated by these gaps, cosmologists have been exploring alternative cosmological models in addition to seeking systematic errors~\cite{DiValentino:2017oaw, FrancoAbellan:2020xnr, RoyChoudhury:2020dmd, He:2023dbn, Tanimura:2023bkh, Caramete:2013bua,Giare:2024akf,Feng:2019jqa,Adil:2023exv,Nesseris:2017vor,Giare:2024syw,deAraujo:2021cnd,Aboubrahim:2024spa,Uzan:2023dsk,Adi:2020qqf,Lynch:2024hzh,Toda:2024ncp,Escamilla:2024xmz,Akarsu:2024eoo,Vagnozzi:2023nrq,Schoneberg:2024ynd,Pedrotti:2024kpn,Chatrchyan:2024xjj,Jiang:2024xnu,Schirra:2024rjq,Toda:2024ncp,Banik:2024vbz,Tiwari:2023jle} (see also these Reviews~\cite{DiValentino:2021izs,Abdalla:2022yfr} and references therein). Within this context, we can highlight an alternative known in the literature as Interacting Dark Energy/Dark Matter (IDE) scenario (see~\cite{Wang:2024vmw} for a review), a model that retains the simplicity of $\Lambda$CDM while introducing an additional parameter, known as the coupling parameter ($\xi$), which governs the non-gravitational interactions between dark matter (DM) and dark energy (DE)~\cite{Kumar:2016zpg, Murgia:2016ccp, Kumar:2017dnp, DiValentino:2017iww,Kumar:2021eev,Pan:2023mie,Benisty:2024lmj,Yang:2020uga,Forconi:2023hsj,Pourtsidou:2016ico,DiValentino:2020vnx,DiValentino:2020leo,Nunes:2021zzi,Yang:2018uae,vonMarttens:2019ixw,Lucca:2020zjb,Zhai:2023yny,Bernui:2023byc,Hoerning:2023hks,Giare:2024ytc,Escamilla:2023shf,SantanaJunior:2024cug,vanderWesthuizen:2023hcl,Silva:2024ift,DiValentino:2019ffd,Li:2024qso,Pooya:2024wsq,Halder:2024uao,Castello:2023zjr,Yao:2023jau,Mishra:2023ueo,Nunes:2016dlj}.

As disagreements and tensions between the $\Lambda$CDM model and observations continue to grow, there is an increasing interest in developing models that extend beyond $\Lambda$CDM, particularly those that incorporate late-time transitions. For example, modifications of gravity that account for late transitions have been explored in~\cite{Alestas:2022xxm, Khosravi:2021csn}. Additionally, transitions involving phantom and quintessence fields have been examined in~\cite{Theodoropoulos:2021hkk, DiValentino:2019exe, Alestas:2021luu, Alestas:2021xes, Alestas:2020zol}. Another avenue of research includes models with a sign-switching cosmological constant, as discussed in~\cite{Akarsu:2019hmw, Akarsu:2022typ, Akarsu:2021fol, Akarsu:2024eoo, Giani:2024nnv,Akarsu:2024qsi, Akarsu:2023mfb,Yadav:2024duq}. Other proposals for late-time transitions have also been investigated, including those in~\cite{Benevento:2020fev, Liu:2024vlt, Frion:2023xwq, Ruchika:2024ymt}.

Recent work by~\cite{Sabogal:2024yha} has highlighted that IDE models can help to alleviate the $H_0$ tension when the DM–DE coupling parameter, $\xi$, is negative, and the $S_8$ tension when $\xi$ is positive. This dual capability suggests that a phenomenological IDE model, which permits both positive and negative values of $\xi$ via distinct interaction channels, has the potential to resolve or mitigate both tensions simultaneously. Motivated by this possibility, in this \textit{paper} we propose a novel interaction model that parametrizes a simple transition in which the direction of energy-momentum transfer reverses after the redshift of equality of DM and DE densities ($z_{\rm eq, dark}$). We develop this framework and assess its capacity to address both the $H_0$ and $S_8$ tensions concurrently.

This paper is structured as follows: Sec.~\ref{model} introduces the IDE model we are considering here. Sec.~\ref{data} describes the methodology and datasets used in this study. The results of the analyses are presented in Sec.~\ref{results}. Finally, Sec.~\ref{conclusions} presents the conclusions drawn from our analyses.

\section{A new dark sector interaction model}
\label{model}
To start reviewing the basic features of IDE models, we adopt a spatially flat Friedmann-Lemaître-Robertson-Walker (FLRW) metric. In the absence of interactions between the fluids associated with DE and DM, their stress-energy tensors, denoted by $T^{\mu\nu}_{\rm x}$ for DE and $T^{\mu\nu}_{\rm c}$ for DM, respectively, are individually covariantly conserved. In IDE models, a parameterization is introduced in the conservation equations such that the individual stress-energy tensors are no longer conserved, but their sum remains conserved~\cite{Wang:2024vmw}. Therefore, the evolution of the covariant derivatives of the stress-energy tensors for DE and DM can be expressed as:

\begin{eqnarray}
\label{eq:covariant}
\sum_{j} \nabla_\mu T_{j}^{\mu\nu} = 0, \hspace{0.5cm}   \, \nabla_\mu T_{j}^{\mu\nu}= \frac{Q_{j} u_{\rm c}^\nu}{a},
\end{eqnarray}\\
\noindent where the index $j$ runs over DE and DM, $a$ is the scale factor, $u_{\rm c}^\nu$ denotes the DM four-velocity vector, and $Q_{\rm c} = -Q_{\rm x} = Q$ is the DE-DM interaction rate with units of energy per volume per time. In the presence of such interaction, the continuity equations for the DM and DE energy densities $\rho_{\rm c}$ and $\rho_{\rm x}$ are modified to:
\begin{eqnarray}
\dot{\rho}_{\rm c}+3 \mathcal{H} \rho_{\rm c} & = Q \, ,\\
\dot{\rho}_{\rm x}+3 \mathcal{H}\left(1+w_{\rm x}\right) \rho_{\rm x} & = -Q \, ,
\end{eqnarray}
in which ${\cal H}$ denotes the conformal Hubble rate, and $w_{\rm x}$ the DE equation of state (EoS). Due to the unknown nature of the dark sector, one must phenomenologically select the functional form of $Q$. In this study, we introduce a variation of a widely recognized interaction kernel~\cite{Gavela:2010tm,DiValentino:2019ffd,DiValentino:2019jae,Zhai_2023}, $Q = \xi(a) {\cal H} \rho_{\rm x}$, where $\xi(a)$ is a new function that governs the strength of the interaction between DE and DM, given by, 
\begin{equation}
    \xi(a)= \xi_{i} \operatorname{sgn}\left[a_{\rm eq,dark}-a\right] \, ,
    \label{xi}
\end{equation}
where the dimensionless free parameter $\xi_{i} \in \mathbb{R}$ characterizes the magnitude and initial direction of the energy-momentum flow when $a \rightarrow 0$. 
The new solutions for the density evolution of dark fluids are now given by

\begin{equation}
\begin{aligned}
\rho_{\rm c} & =\rho_{\rm c, 0} \, a^{-3} + \rho_{\rm x, 0}\, a^{-3} \left( \frac{\xi_{i}}{3 w_{\rm x}+\xi_{i}} \right) \\
& \times \left[ \frac{\xi_{i} + 3 w_{\rm x} \left(1 - 2 a_{\rm eq}^{\xi_{i} - 3 w_{\rm x}} \right) }{\xi_{i}-3 w_{\rm x}} -a_{\rm eq}^{2 \xi_{i}} a^{-3(w_{\rm x}+\xi_{i}/3)} \right] \\
\rho_{\rm x} & =\rho_{\rm x, 0}\, a^{-3\left(1+w_{\rm x} + \xi_{i}/3\right)} \, a_{\rm eq}^{2\xi_{i}}
\end{aligned}
\end{equation}

Note that, due to the transition in the interaction between the dark components, the density evolutions in these models differ from the standard cases~\cite{Kumar:2016zpg, Murgia:2016ccp, Benisty:2024lmj,Giare:2024smz,Silva:2024ift,Sabogal:2024yha}. The usual solutions are recovered when $a_{\rm eq,dark} = 1$ (for simplicity in the above equations, $a_{\rm eq,dark}$ was simplified to $a_{\rm eq}$), indicating no transition in the energy-momentum transfer between the fluids.

The sign-switching transition of $\xi$ is implemented via the signum function (sgn) at $a_{\rm eq,dark}$, corresponding to the cosmic epoch when the total matter and dark energy densities are equal, $\Omega_{\rm m}/\Omega_{\rm x}(a_{\rm eq,dark}) = 1$,
\begin{equation}
\label{a_t}
 a_{\rm eq,dark} = \dfrac{1}{1+z_{\rm eq,dark}} \equiv \left[ \dfrac{ \left( 1 + \dfrac{\xi_{i}}{3w_{\rm eff}} \right) }{ \left(\dfrac{\Omega_{{\rm m},0}}{\Omega_{{\rm x},0}} + \dfrac{\xi_{i}}{3w_{\rm eff}} \right)} \right]^{\dfrac{1}{3w_{\rm eff}}} \, ,
\end{equation}
with the effective equation of state denoted as $w_{\rm eff} = w_{\rm x} + \xi_{i}/3$.
 
In this parametrization, when $\xi < 0$, energy-momentum transfers from DM to DE, while for $\xi > 0$, energy flows from DE to DM. It should be noted that in our formalism, $\Lambda$CDM is recovered for $\xi_{i}=0$.

The choice of redshift for the transition in dark coupling is motivated by the cosmic coincidence problem~\cite{Chimento:2003iea,Velten:2014nra,Zimdahl:2002zb}, which occurs when the densities of DM and DE are exactly the same. Cosmic coincidence refers to the observation that, at the present epoch of the universe's evolution, the energy densities associated with DM and DE are of similar magnitudes, leading to their comparable effects on the dynamics of the universe. The model presented here is not designed to solve the cosmic coincidence problem; rather, we use this fact to propose a transition in the sign change of interaction within the dark sector. On the other hand, it has been investigated whether a sign-change transition in the dark sector yields stable predictions or is even possible in model-independent analyses~\cite{Halder:2024uao, Bonilla:2021dql, Escamilla:2023shf,DiGennaro:2022ykp,Oikonomou:2024zhs}.

\begin{figure}[htpb!]
   \centering
    \includegraphics[width=0.5\textwidth]{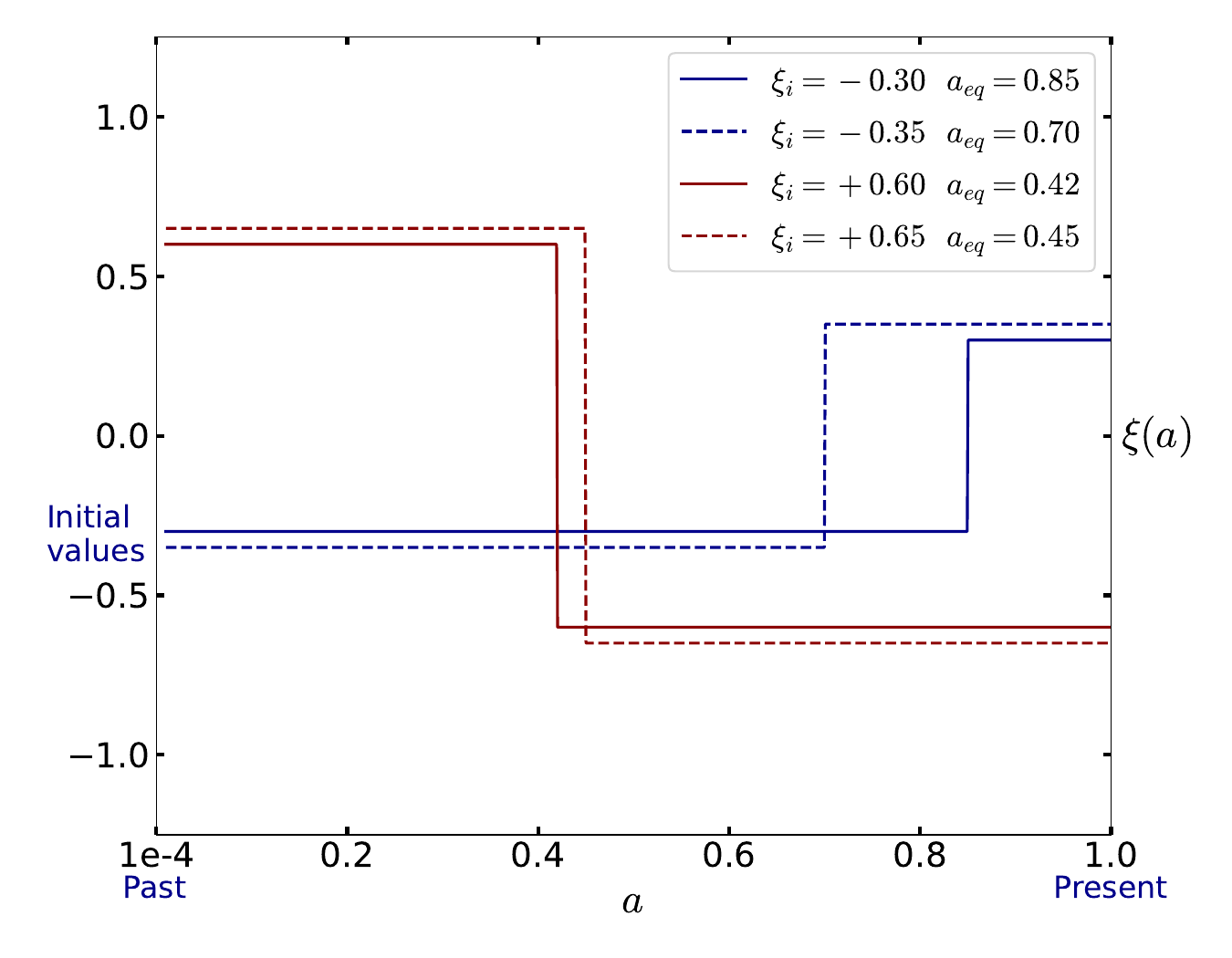} 
    \caption{A visual representation of our model demonstrating the shift in the direction of the energy-momentum flow between DE and DM.}
    \label{fig:model}
\end{figure}

Since the dark interaction is assumed to always exist throughout cosmic evolution, we will use the following strategies in our analyses to determine the initial signal and when the signal will change:

\begin{itemize}
  \item Step 1: Prior to the transition described by Eq. (\ref{a_t}), the sign of $\xi_i$ remains free, to be determined solely by fitting the data during the MCMC analysis.
  
  \item Step 2: To prevent instabilities, we adopt the standard procedure, where the quantities $(1 + w_{\rm x})$ and $\xi$ must have opposite signs~\cite{Gavela:2009cy, He:2008si}. Accordingly, depending on the value of $\xi_{i}$, we assign $w_{\rm x} = -0.999$ for $\xi_{i} < 0$ or $w_{\rm x} = -1.001$ for $\xi_{i} > 0$.
  
  \item Step 3: Once the initial value of $\xi_i$ is obtained, its sign is reversed at $a_{\rm eq,dark}$. Following this transition, the direction of energy-momentum transfer is also inverted.

  \item Step 4: The magnitude of the interaction remains constant before and after the transition. In other words, the magnitude of $\xi$ does not change; only its sign reverses at the transition.

\end{itemize}

The steps outlined above provide an efficient strategy for implementing this new model. Our proposal builds on well-established models (see~\cite{Gavela:2010tm, Nunes:2022bhn, Gavela:2009cy, He:2008si, DiValentino:2017iww, DiValentino:2019ffd, Zhai:2023yny, Gariazzo:2021qtg, Bernui:2023byc}), now modified to effectively explore the possibility of sign switching in dark sector coupling interactions across the entire parameter space. Naturally, more complex scenarios could be proposed and explored. However, in this work, we focus on a simpler case to facilitate an observational investigation of the new scenario. For instance, if the best-fit value of the coupling is $\xi_i < 0$ when $a \rightarrow 0$, then according to our model, $\xi$ must necessarily become positive after the transition, i.e., $\xi > 0$. The magnitude of $\xi$ will remain the same as $\xi_i$, with only its sign changing. Specifically, for all cosmic times prior to the transition, we have $\xi < 0$, and after the transition, $\xi > 0$, while the magnitude of $\xi$ remains constant throughout cosmic history. The interpretation is reversed if $\xi_i > 0$. Figure~\ref{fig:model} visually illustrates the reversal of energy-momentum flow for various transition points and coupling magnitudes.

Within the framework of linear perturbations, as already well known in the literature, the non-gravitational interaction between DE and DM also affects their evolution. In this case, the solution remains the same as in the standard IDE model~\cite{Gavela:2010tm, Nunes:2022bhn, Gavela:2009cy, He:2008si}, with a simple transformation of the coupling parameter, $\xi \rightarrow \xi(a)$. However, this does not imply that the effects are identical. The evolution of perturbations is also shaped by variations in the expansion rate, ${\cal H}$, as well as the modified solutions for $\rho_{\rm c}$ and $\rho_{\rm x}$. In the scenario explored in this work, the Boltzmann equations take the following form in the synchronous gauge

\begin{subequations}
\begin{eqnarray}
\dot{\delta}_{\rm c} & = & - \left[ \theta_{\rm c} + \frac{\dot{h}}{2} \right] + \frac{\rho_{\rm x}}{\rho_{\rm c}} \Xi(a) \\
& & + \, \Gamma(a) \left[\frac{3 \mathcal{H}}{k^2}  (w_{\rm x}-1) +  \left(\delta_{\rm x} - \delta_{\rm c}\right)  \right]\, , \nonumber \\
\dot{\theta}_{\rm c} & = & -\mathcal{H} \theta_{\rm c} \, , \\ 
\dot{\delta}_{\rm x} & = & -\left(1 + w_{\rm x}\right)\left[ \theta_{\rm x} +  \frac{\dot{h}}{2}\right] - \Xi(a) \\
&& - \dfrac{3\mathcal{H}}{k^{2}} \left(1 - w_{\rm x}\right) \left[ k^{2} \delta_{\rm x} + 3 \mathcal{H} \theta_{\rm x} \right]\,, \nonumber \\ 
\dot{\theta}_{\rm x} & = & 2 \mathcal{H} \theta_{\rm x} + \frac{k^2}{1 + w_{\rm x}} \delta_{\rm x} + \Gamma(a) \frac{\rho_{\rm c} \left( 2\theta_{\rm x} - \theta_{\rm c} \right) }{\rho_{\rm x}(1 + w_{\rm x})}\,, 
\label{EB_4}
\end{eqnarray}
\label{E_boltzmann}
\end{subequations}
where the source terms of interaction are given by
\begin{subequations}
\begin{eqnarray}
\Gamma & = & \xi_{i} \mathcal{H} \frac{\rho_{\rm x}}{\rho_{\rm c}} \operatorname{sgn}\left[a_{\rm eq}-a\right]\,, \\
\Xi & = &  \xi_{i} \left[ \frac{3 \mathcal{H}^{2}}{k^2}(1 - w_{\rm x}) + \frac{k v_T}{3} + \frac{\dot{h}}{6}\right] \operatorname{sgn}\left[a_{\rm eq}-a\right].\,\,\,\,\,
\end{eqnarray}
\label{interaction_terms}
\end{subequations}

In the absence of any interaction within the dark sector, the standard $\Lambda$CDM model is recovered. In what follows, we will discuss our statistical methodology and the dataset used in our analysis.

\section{Datasets and methodology}
\label{data}
To test our new theoretical framework, we implemented it in the Boltzmann solver code \texttt{CLASS}~\cite{Blas:2011rf} and used the publicly available sampler \texttt{MontePython}~\cite{Brinckmann:2018cvx, Audren:2012wb} to perform Markov Chain Monte Carlo (MCMC) analyses, ensuring a Gelman-Rubin convergence criterion~\cite{Gelman_1992} of $R-1 \leq 10^{-2}$ in all runs. We assumed flat priors on the set of sampled cosmological parameters \{$\Omega_{\rm b} h^2$, $\Omega_{\rm c} h^2$, $\tau_{\rm reio}$, $100\theta_{\mathrm{s}}$, $\log(10^{10} A_{\mathrm{s}})$, $n_{\mathrm{s}}$, $\xi_{i}$\}, where the first six are baseline parameters within the $\Lambda$CDM context. Specifically, these parameters include the present-day physical density parameters of baryons ($\omega_b = \Omega_{\rm b} h^2$) and dark matter ($\omega_c = \Omega_{\rm c} h^2$), the optical depth of reionization ($\tau_{\rm reio}$), the angular size of the sound horizon at recombination ($\theta_{\mathrm{s}}$), the amplitude of the primordial scalar perturbation ($A_{\mathrm{s}}$), and the scalar spectral index ($n_{\mathrm{s}}$). The ranges of the priors are: $\omega_b \in [0.0, 1.0]$, $\omega_{\text{cdm}} \in [0.0, 1.0]$, $100 \theta_s \in [0.5, 2.0]$, $\ln(10^{10} A_s) \in [1.0, 5.0]$, $n_s \in [0.1, 2.0]$, $\tau_{\text{reio}} \in [0.004, 0.8]$, and $\xi_{i} \in [-1.5, 1.5]$. In all analyses, we used the Python package \texttt{GetDist}\footnote{\url{https://github.com/cmbant/getdist}} to analyze the MCMC chains and extract the numerical results, as well as the 1D posteriors and 2D marginalized probability contours. \footnote{In all analyses carried out in this paper, we expanded the prior ranges by up to one order of magnitude beyond the values mentioned and those typically used in standard analyses. We did not find any significant differences in the results, indicating that our conclusions are robust to the choice of prior ranges. }

\begin{table*}[htpb!]
\begin{center}
\caption{Marginalized constraints, mean values with $68\%$ CL, on the free and some derived parameters of the new IDE model from the CMB dataset and its combinations with DESI, PPS, PP, Union3, and DESY5. In the last rows, we provide $\Delta \chi^2_{\text{min}} = \chi^2_{\text{min (IDE)}} - \chi^2_{\text{min ($\Lambda$CDM)}}$, $\Delta \text{AIC} = \text{AIC}_{\text{IDE}} - \text{AIC}_{\text{$\Lambda$CDM}}$, and the Bayes factors $\ln{\mathcal{B}_{ij}}$ as defined in Eq. (\ref{BayesFactor}). Negative values of \( \Delta \chi^2_{\text{min}} \) and \( \Delta \text{AIC} \) indicate a better fit of the IDE model compared to the $\Lambda$CDM model, while a negative value of \( \ln{\mathcal{B}_{ij}} \) indicates a preference for the IDE model over the $\Lambda$CDM model. }
\label{tab_results_I}
\renewcommand{\arraystretch}{1.5}
\resizebox{\textwidth}{!}{
\begin{tabular}{l||cccc||ccc} 
\hline
\textbf{Parameter} & \textbf{CMB} & \textbf{CMB + DESI} & \textbf{CMB + PPS} & \textbf{CMB + PPS + DESI} & \textbf{CMB + PP} & \textbf{CMB + Union3} & \textbf{CMB + DESY5} \\ 
\hline\hline
$100\Omega_{\rm b} h^2$ & 
$2.242\pm 0.015$ & 
$2.246\pm 0.015$ & 
$2.251\pm 0.014$  & 
$2.254\pm 0.014$ & 
$2.237 \pm 0.014$ & 
$2.236\pm 0.015$  & 
$2.235^{+0.013}_{-0.015}$ \\

$\Omega_{\rm c} h^2 $ & $0.1271^{+0.0042}_{-0.0089}$ & $0.1243^{+0.0033}_{-0.0030}$ &
$0.1241^{+0.0021}_{-0.0018}$ & $0.1235^{+0.0021}_{-0.0016}$ & $0.1200^{+0.0020}_{-0.0017}$ & $0.1217^{+0.0018}_{-0.0038}$  & $0.1240^{+0.0022}_{-0.0048}$\\

$100\theta_\mathrm{s}$  & $1.04192\pm 0.00030$ & $1.04196\pm 0.00028$ & $1.04203\pm 0.00029$ & $1.04209\pm 0.00028$ & $1.04186\pm 0.00029$ & $1.04186\pm 0.00030$ & $1.04185\pm 0.00029$\\

$\tau_\mathrm{reio}$ & $0.0540^{+0.0066}_{-0.0078}$ & $0.0554\pm 0.0080$ &
$0.0567^{+0.0066}_{-0.0077}$ & $0.0575^{+0.0065}_{-0.0076}$ & $0.0549^{+0.0064}_{-0.0078}$ & $0.0538\pm 0.0074$ & $0.0548^{+0.0066}_{-0.0074}$  \\

$n_\mathrm{s}$ & $0.9670\pm 0.0046$ & 
$0.9681\pm 0.0041$ &
$0.9689\pm 0.0040$ & 
$0.9695\pm 0.0037$ & 
$0.9657\pm 0.0040$ &
$0.9649\pm 0.0042$ & 
$0.9645\pm 0.0041$\\

$\log(10^{10} A_\mathrm{s})$ & 
$3.046\pm 0.016$ &
$0.9681\pm 0.0041$ & 
$3.047^{+0.013}_{-0.015}$ & 
$3.049\pm 0.014$ & 
$3.046^{+0.013}_{-0.015}$  
& $3.045\pm 0.014$  
& $3.047\pm 0.014$\\

$\xi_{i}$ & $-0.43^{+0.72}_{-0.49}$ &
$-0.37\pm 0.20$ &
$-0.35\pm 0.11$ & 
$-0.32\pm 0.10$ & 
$0.12^{+0.14}_{-0.12}$  & $0.22^{+0.17}_{-0.11}$  & $0.325\pm 0.093$ \\\hline

$H_0$ (km/s/Mpc) & $70.4^{+3.3}_{-5.2}$ &
$70.3\pm 1.1$ & 
$70.34\pm 0.71$ & 
$70.29\pm 0.58$ & 
$66.45\pm 0.93$  & $65.6^{+1.1}_{-1.3}$  & $64.69^{+0.87}_{-0.71}$ \\

$\Omega_{\rm m }$ & $0.304^{+0.016}_{-0.030}$ & $0.2986^{+0.0039}_{-0.0051}$ & 
$0.2976\pm 0.0045$ & 
$0.2969\pm 0.0035$ & 
$0.3241^{+0.0081}_{-0.013}$ & $0.337^{+0.013}_{-0.023}$ & $0.352^{+0.011}_{-0.022}$\\

$S_8$ & $0.769^{+0.099}_{-0.049}$ &
$0.773^{+0.022}_{-0.025}$ &
$0.771\pm 0.016$ &
$0.772\pm 0.013$ &
$0.845^{+0.017}_{-0.011}$ & $0.852^{+0.015}_{-0.0086}$ & $0.857^{+0.011}_{-0.0086}$ \\ 

$z_{\rm eq, dark}$ & $0.248^{+0.069}_{-0.088}$ & $0.248^{+0.022}_{-0.038}$ &
$0.249^{+0.017}_{-0.024}$ & 
$0.257^{+0.015}_{-0.024}$ & 
$0.320^{+0.029}_{-0.038}$ & 
$0.321^{+0.031}_{-0.038}$ & 
$0.321\pm 0.042$\\ 
 \hline
$\Delta \chi^{2}_{\mathrm{min}}$ & $\quad 0.02$ & $-3.28$ & $-11.36$ & $-10.50$ &$ -0.12$ & $-0.96$ & $-4.56$\\
$\Delta \mathrm{AIC}$ & $\quad 2.02$ & $-1.28$ & $-9.36$ & $-8.50$ &$\quad 1.88$ & $\quad 1.04$ & $-2.56$\\
$\ln \mathcal B_{ij}$ & $-0.15$ & $-0.74$ & $-2.80$ & $-2.84$ &$\quad 1.29$ & $\quad 0.38$ & $-1.26$\\
\hline \hline
\end{tabular}}
\end{center}
\end{table*}

\begin{figure*}[htpb!]
    \centering
    \includegraphics[width=0.48\textwidth]{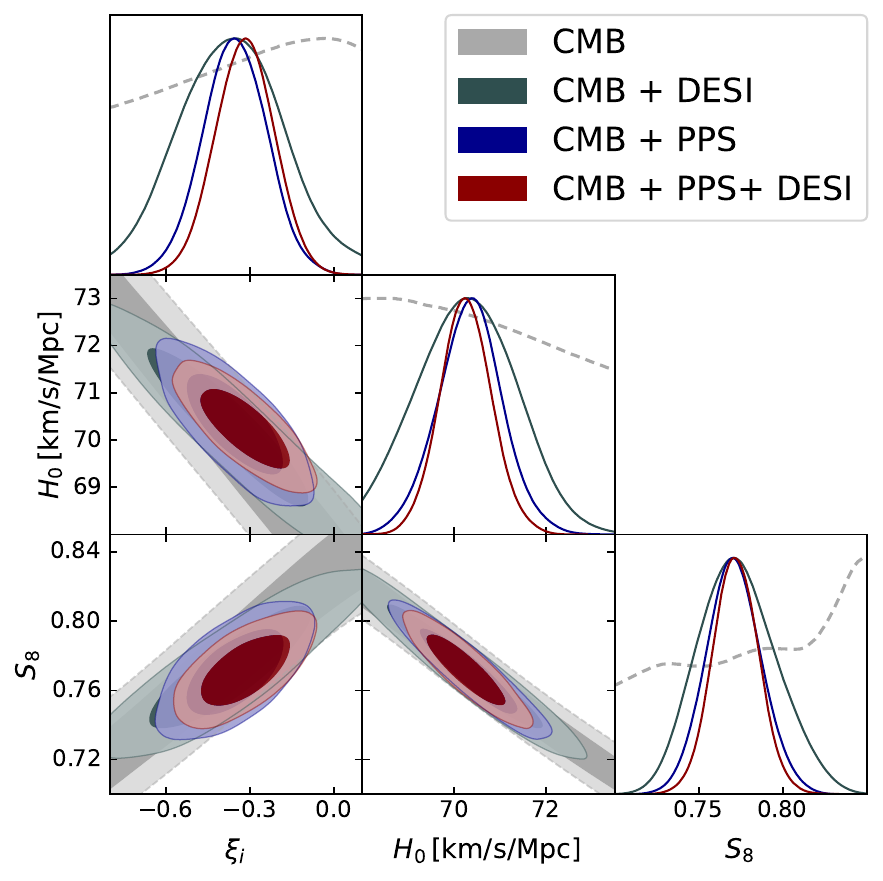} \hspace{0.02\textwidth}
    \includegraphics[width=0.48\textwidth]{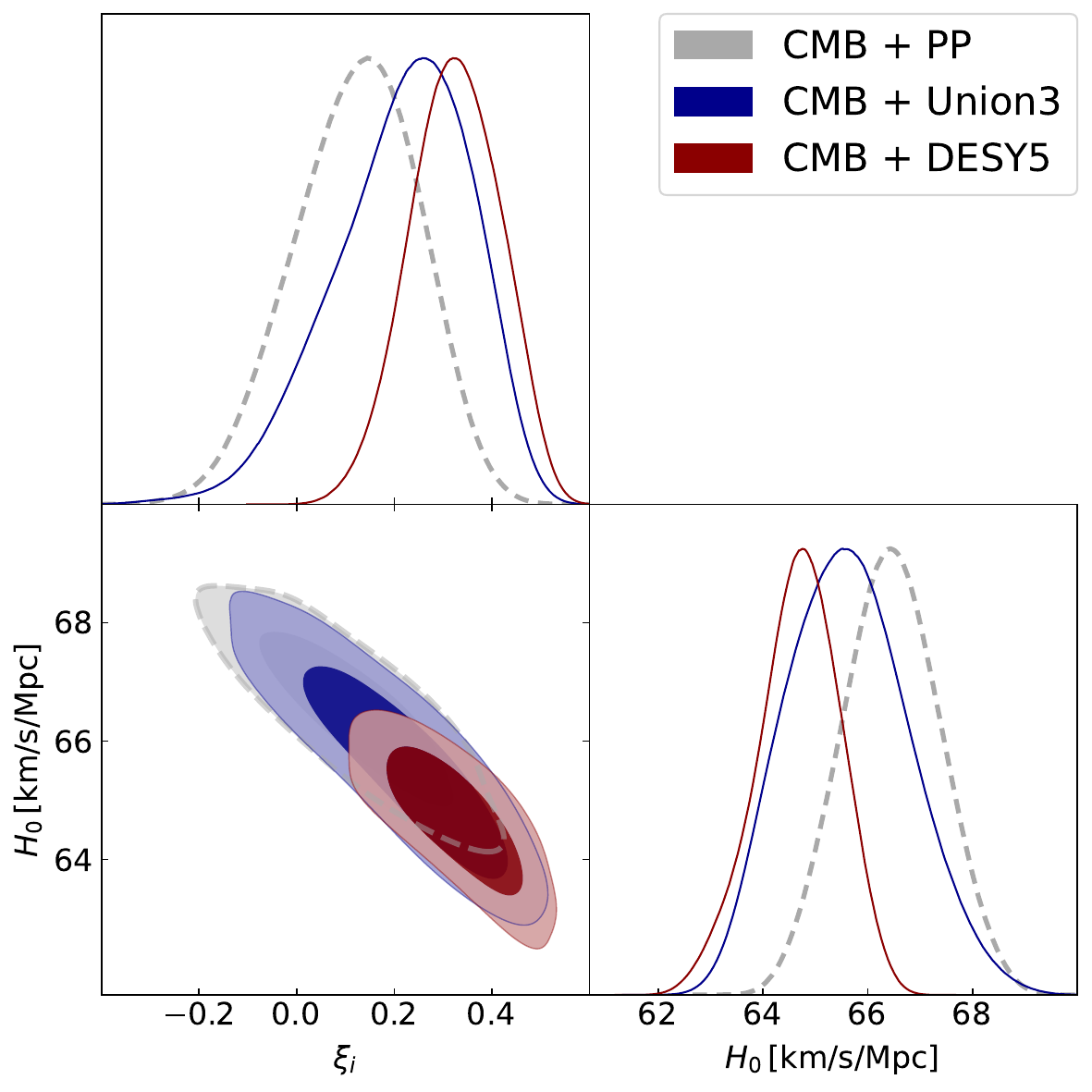}
    \caption{\textbf{Left panel:} Marginalized one-dimensional posterior distributions and contours (68\% and 95\% CL) for the parameters $\xi_i$, $H_{0}$, and $S_{8}$ in the IDE model from CMB and its combinations with DESI and PPS data, as shown in the legend. \textbf{Right panel:} Same as the left panel but for the parameters $\xi_i$ and $H_0$ from combinations of CMB with PP, Union3, and DESY5, as described in the legend.}
    \label{fig:cmb_xi}
\end{figure*}

The datasets used in the analyses are described below.
\begin{itemize}

\item \textit{Cosmic Microwave Background} (\textbf{CMB}): We utilize temperature and polarization anisotropy measurements of the CMB power spectra from the Planck satellite~\cite{Planck:2018vyg}, along with their cross-spectra from the Planck 2018 legacy data release. Specifically, we employ the high-$\ell$ \texttt{Plik} likelihood for TT (in the multipole range $30 \leq \ell \leq 2508$), TE, and EE ($30 \leq \ell \leq 1996$), as well as the low-$\ell$ TT-only ($2 \leq \ell \leq 29$) likelihood and the low-$\ell$ EE-only ($2 \leq \ell \leq 29$) \texttt{SimAll} likelihood~\cite{Planck:2019nip}. Additionally, we include CMB lensing measurements, which are reconstructed from the temperature 4-point correlation function~\cite{Planck:2018lbu}. We refer to this dataset as \texttt{CMB}.

\item \textit{Baryon Acoustic Oscillations} (\textbf{DESI}): We consider Baryon Acoustic Oscillation (BAO) measurements from the Dark Energy Spectroscopic Instrument (DESI) survey, which encompasses observations of galaxies and quasars~\cite{DESI:2024uvr} as well as Lyman-$\alpha$~\cite{DESI:2024lzq} tracers. These measurements, summarized in Table I of Ref.~\cite{DESI:2024mwx}, cover both isotropic and anisotropic BAO measurements within the redshift range $0.1 < z < 4.2$, divided into seven redshift bins. The isotropic BAO measurements are expressed as $D_{\mathrm{V}}(z)/r_{\mathrm{d}}$, where $D_{\mathrm{V}}$ represents the angle-averaged distance, normalized to the comoving sound horizon at the drag epoch. The anisotropic BAO measurements include $D_{\mathrm{M}}(z)/r_{\mathrm{d}}$ and $D_{\mathrm{H}}(z)/r_{\mathrm{d}}$, where $D_{\mathrm{M}}$ denotes the comoving angular diameter distance and $D_{\mathrm{H}}$ represents the Hubble horizon. We also account for the correlation between measurements of $D_{\mathrm{M}}/r_{\mathrm{d}}$ and $D_{\mathrm{V}}/r_{\mathrm{d}}$. We refer to this dataset as \texttt{DESI}.

\item \textit{Type Ia Supernovae} (\textbf{SN Ia}): Type Ia supernovae act as standardizable candles, providing a crucial method for measuring the universe's expansion history. Historically, SN Ia played a pivotal role in the discovery of the accelerating expansion of the universe~\cite{SupernovaSearchTeam:1998fmf, SupernovaCosmologyProject:1998vns}, building upon earlier, more complex arguments supporting $\Lambda$-dominated models from large-scale structure observations. In this work, we will use the following recent samples:
\begin{enumerate}
    \item [(i)] \textbf{PantheonPlus and PantheonPlus\&SH0ES}: We integrated the latest distance modulus measurements from SN Ia in the PantheonPlus sample~\cite{pantheonplus} which includes 1701 light curves from 1550 distinct SN Ia events, spanning a redshift range of $0.01$ to $2.26$. We designate this dataset as \texttt{PP}. Additionally, we consider a version of this sample that uses the latest SH0ES Cepheid host distance anchors~\cite{Riess:2021jrx} to calibrate the absolute magnitude of SN Ia, rather than centering a prior on the $H_{0}$ value from SH0ES. This approach allows for more robust results, and this version of the dataset is referred to as \texttt{PPS}. 

    \item [(ii)] \textbf{Union 3.0}: The Union 3.0 compilation, consisting of 2087 SN Ia within the range $0.001 < z < 2.260$, was presented in~\cite{Rubin:2023ovl}. Notably, 1363 of these SN Ia overlap with the PantheonPlus sample. This dataset features a distinct treatment of systematic errors and uncertainties, employing Bayesian Hierarchical Modeling. We refer to this dataset as \texttt{Union3}.

    \item [(iii)] \textbf{DESY5}: As part of their Year 5 data release, the Dark Energy Survey (DES) recently published results from a new, homogeneously selected sample of 1635 photometrically-classified SN Ia with redshifts spanning $0.1 < z < 1.3$~\cite{DES:2024tys}. This sample is complemented by 194 low-redshift SN Ia (shared with the PantheonPlus sample) in the range $0.025 < z < 0.1$. We refer to this dataset as \texttt{DESY5}.
\end{enumerate}
\end{itemize}

\subsection{Tension metrics}

To quantify the degree of tension between two datasets, we will use the quadratic estimator introduced in ~\cite{Addison:2015wyg}. This estimator is defined as:
\begin{equation}
    \chi^2 = \left(\mathbf{x}_i - \mathbf{x}_j\right)^\mathrm{T} \left(\mathcal{C}_i + \mathcal{C}_j\right)^{-1} \left(\mathbf{x}_i - \mathbf{x}_j\right),
    \label{N-tension}
\end{equation}
where $ \mathbf{x}_i $ and $ \mathbf{x}_j $ are vectors containing the mean values of the cosmological parameters inferred from datasets $ i $ and $ j $, respectively. The quantities $ \mathcal{C}_i $ and $ \mathcal{C}_j $ denote the corresponding covariance matrices associated with the parameter estimates for each dataset.

This estimator provides a robust statistical measure of the agreement or tension between the datasets, reflecting how well they align within the parameter space of the theoretical model. By comparing the datasets in terms of their mean values and associated uncertainties, the method helps to identify discrepancies that may point to systematic errors, inconsistencies, or the need for modifications to the underlying theoretical framework.

\subsection{Bayesian Model Comparison}

To better assess the level of agreement (or disagreement) between the models and their association with each analyzed dataset, we perform a statistical comparison of the IDE model with the $\Lambda$CDM scenario using the well-known Akaike Information Criterion (AIC)~\cite{Akaike:1974vps}.

The AIC is defined as:
\begin{equation}
\mathrm{AIC} \equiv -2 \ln \mathcal{L}_{\text{max}} + 2N,
\label{AIC}
\end{equation}
where $\mathcal{L}_{\text{max}}$ is the maximum likelihood of the model, and $N$ is the total number of free parameters in the model. The lower the value of AIC, the better the model's performance in terms of both the quality of the fit and its complexity. Specifically, the AIC incorporates a penalty for the number of parameters in the model, discouraging the inclusion of unnecessary parameters that could lead to overfitting.

Beyond a simple AIC assessment, the Bayesian Evidence (BE) offers a more formal quantification by calculating the Bayes factor, which compares an extended model relative to a baseline model. The BE accounts for the trade-off between the higher likelihood of the extended model and the penalty imposed by its increased complexity and prior volume. To illustrate the BE calculation, consider a dataset $\mathbf{x}$ and two competing models, $\mathcal{M}_i$ and $\mathcal{M}_j$, characterized by parameters $\boldsymbol{\theta}_i$ and $\boldsymbol{\theta}_j$, respectively. In the case of nested models, $\boldsymbol{\theta}_i$ may be a subset of $\boldsymbol{\theta}_j$, or vice versa. The Bayes factor $\mathcal{B}_{ij}$, which quantifies the strength of evidence in favor of model $\mathcal{M}_i$ over model $\mathcal{M}_j$, is given (under the assumption of equal prior probabilities for the two models, which is often the case) by:

\begin{equation}
\mathcal{B}_{ij} = \frac{p(\mathcal{M}_i|\mathbf{x})}{p(\mathcal{M}_j|\mathbf{x})} = \frac{\displaystyle \int d \boldsymbol{\theta}_i \, \pi\left(\boldsymbol{\theta}_i | \mathcal{M}_i\right) \mathcal{L}\left(\mathbf{x} | \boldsymbol{\theta}_i, \mathcal{M}_i\right)}{\displaystyle\int d \boldsymbol{\theta}_j \, \pi\left(\boldsymbol{\theta}_j | \mathcal{M}_j\right) \mathcal{L}\left(\mathbf{x} | \boldsymbol{\theta}_j, \mathcal{M}_j\right)},
\label{BayesFactor}
\end{equation}
where $p(\mathcal{M}_i|\mathbf{x})$ represents the Bayesian evidence for model $\mathcal{M}_i$, $\pi(\boldsymbol{\theta}_i|\mathcal{M}_i)$ denotes the prior distribution of the model parameters, and $\mathcal{L}(\mathbf{x}|\boldsymbol{\theta}_i, \mathcal{M}_i)$ is the likelihood of the data given the model's parameters. A Bayes factor $\mathcal{B}_{ij} > 1$ indicates that the data provide stronger support for model $\mathcal{M}_i$ (in this case, the $\Lambda$CDM model) over model $\mathcal{M}_j$ (IDE), even if model $\mathcal{M}_j$ demonstrates a superior data fit due to the penalty on the latter model’s increased complexity. The interpretation of different values of $\mathcal{B}_{ij}$ (or equivalently $\ln \mathcal{B}_{ij}$) follows various qualitative scales. In this work, we adopt the scale proposed by Raftery~\cite{Kass:1995loi}, which is summarized as:

\begin{itemize}
    \item $0 \leq |\ln \mathcal{B}_{ij}| < 1$: Weak evidence.
    \item $1 \leq |\ln \mathcal{B}_{ij}| < 3$: Definite/Positive evidence.
    \item $3 \leq |\ln \mathcal{B}_{ij}| < 5$: Strong evidence.
    \item $|\ln \mathcal{B}_{ij}| \geq 5$: Very strong evidence.
\end{itemize}

Therefore, a negative value of $\ln \mathcal{B}_{ij}$ suggests a preference for the IDE model over $\Lambda$CDM, while positive values of $\ln \mathcal{B}_{ij}$ indicate a preference for $\Lambda$CDM over the IDE model. To compute the Bayes factors, we employ the \texttt{MCEvidence} package, which is publicly available~\cite{Heavens:2017hkr,Heavens:2017afc}.\footnote{Can be accessed at the following link: \url{https://github.com/yabebalFantaye/MCEvidence}.}

In the following sections, we present our main findings.

\section{Results and Discussions}
\label{results}

In Table~\ref{tab_results_I}, we summarize the results of our analyses, first considering CMB data alone, followed by \textbf{joint} analyses with PPS and DESI. As is well known for IDE models, we observe that CMB data alone exhibit a degenerate parameter space, meaning they lack strong constraining power when considering dark sector coupling. Since the predicted value of $H_0$ is high and consistent with local measurements, we can, without loss of generality, combine CMB with PPS. This contrasts with the $\Lambda$CDM scenario, where these datasets cannot be statistically consistently combined. Furthermore, we incorporate recent BAO-DESI measurements to improve constraints on the model's baseline.


Compared to previous studies based on the constant coupling $\xi$ (see for instance ~\cite{DiValentino:2017iww, DiValentino:2019ffd, Zhai:2023yny, Gariazzo:2021qtg, Bernui:2023byc, Giare:2024smz}), the main novelty of our model lies in demonstrating that a late-time transition in the coupling sign, from $\xi_i < 0$ to $\xi_i > 0$, has a potentially non-negligible impact, leading to higher $H_{0}$ values. Although the $\xi$-$S_8$ correlation had already been identified in earlier works, particularly in~\cite{Kumar:2016zpg, Murgia:2016ccp}, our analysis shows that this transition not only supports high $H_0$ values, consistent with distance ladder measurements, but also predicts lower $S_8$ values, which are more closely aligned with weak lensing results (see Figure~\ref{fig:cmb_xi}). Furthermore, this model introduces additional improvements, including tighter constraints on $\Omega_m$, now centered around $\sim 0.3$ (see Table~\ref{tab_results_I}), ensuring consistency with observational data and enhancing its ability to effectively probe large-scale structure formation.

At this stage, considering our joint analysis of CMB + PPS + DESI data, we find $S_8 = 0.772 \pm 0.013$ at 68\% CL, with corresponding constraints $H_0 = 70.29 \pm 0.58$ at 68\% CL. In this scenario, the $H_0$ tension is reduced to 2.3$\sigma$, while the constraints on $S_8$ show no tension with the formation and evolution of cosmic structures at late times, as predicted by surveys such as DES~\cite{DES:2021wwk} and KiDS-1000~\cite{KiDS:2020suj}. Since the late-time growth of structure is modified in these models, a more quantitative comparison requires re-analyzing the weak lensing and galaxy clustering measurements from these surveys, as the constraints in this context are also model-dependent. However, we do not expect this to significantly alter the correlations in the $\xi$-$S_8$ plane. In future work, the full DES and/or KiDS-1000 likelihood for IDE models should be developed to confirm our findings. Partially, we can conclude that due to statistical correlations for a model with $\xi_i < 0$, and with a late transition to $\xi_i > 0$, the scenario can generate simultaneously high values of $H_0$ and low values of $S_8$, compared to the $\Lambda$CDM model case.

From the joint CMB + PPS analysis, we observe that $\xi_i < 0$ at more than 3$\sigma$, with a chi-square difference of $\Delta \chi^2_{\text{min}} = -11.36$, indicating a significant improvement over the $\Lambda$CDM model, reinforce by $\Delta \text{AIC} = -9.36$ and a definite evidence of $\ln{\mathcal{B}_{ij}} = -2.8$. A similar trend is observed in the CMB + PPS + DESI analysis, where $\Delta \chi^2_{\text{min}} = -10.5$, $\Delta \text{AIC} = -8.5$ and, $\ln{\mathcal{B}_{ij}} = -2.84$. This provides strong evidence for $\xi_i < 0$ in the global fit compared to the $\Lambda$CDM model. In Table~\ref{tab_results_I}, we also present the constraints on $z_{\rm eq, dark}$, reconstructed as a derived parameter from eq. (\ref{a_t}). We find $z_{\rm eq, dark} \sim 0.25$ across all the analyses discussed above.

It is worth mentioning that the joint CMB + DESI analysis alone suggests evidence for $\xi_i < 0$ at nearly 2$\sigma$, with correspondingly high values of $H_0$ and low values of $S_8$.  
In addition, the analysis yields negative values for the quantities $\Delta \chi^2_{\text{min}} = -3.28$, $\Delta \text{AIC} = -1.28$, and $\ln \mathcal{B}_{ij} = -0.74$, demonstrating a mild preference for the model over $\Lambda$CDM.

As shown in Table~\ref{tab_results_I}, the analysis using only CMB data is effectively neutral, indicating that the $\Lambda$CDM and IDE models are statistically indistinguishable. This suggests that the IDE model does not degrade the fit to the CMB data. However, as discussed in the previous paragraphs, incorporating additional datasets in combination with the CMB (e.g., CMB + DESI, CMB + PPS, CMB + PPS + DESI, and CMB + DESY5) can, in some cases—primarily due to the inclusion of PPS samples—reveal a significant preference for the IDE model.
\\


\begin{figure*}[htpb!]
    \centering
    \includegraphics[width=0.46\textwidth]{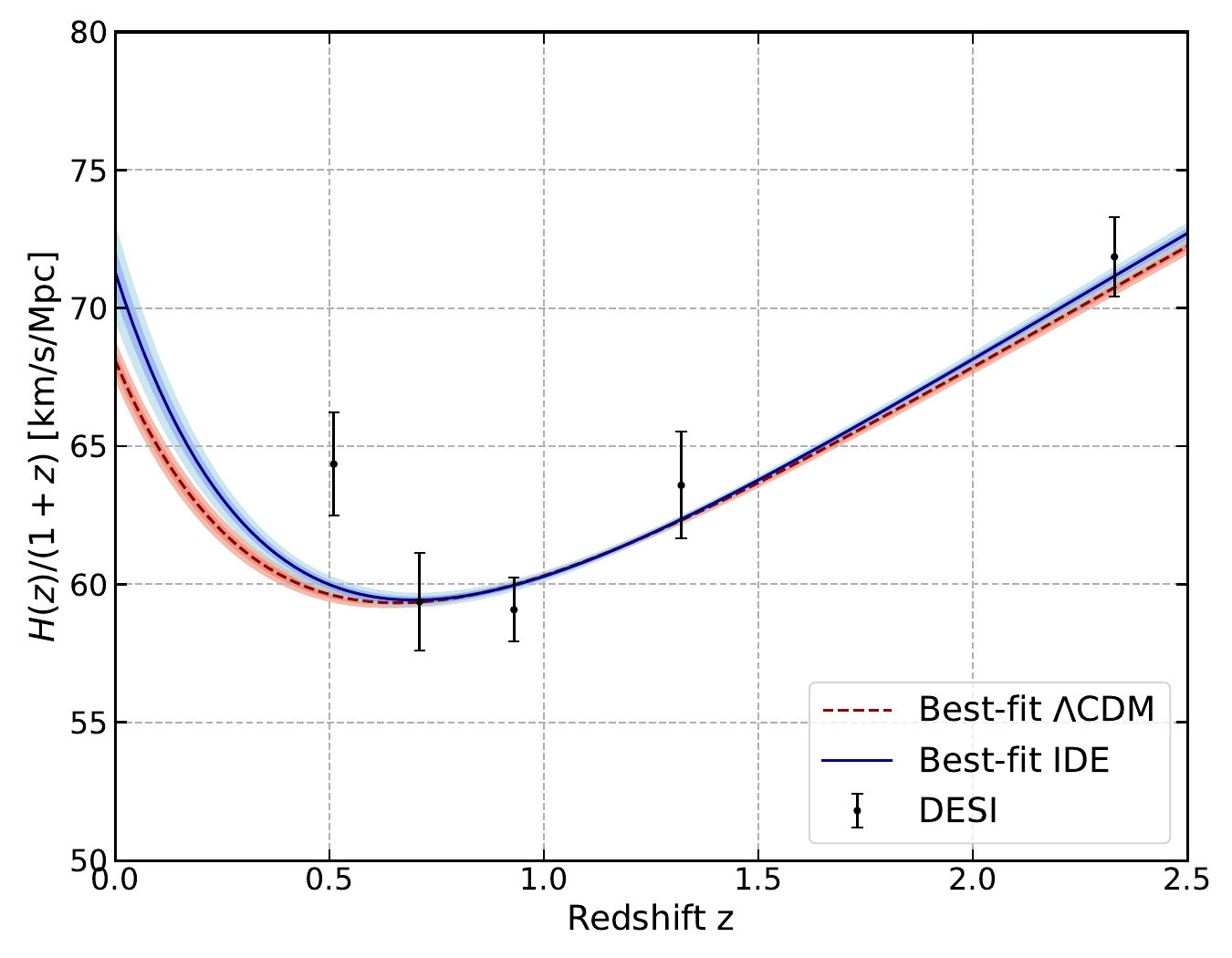} \,\,\,
    \includegraphics[width=0.48\textwidth]{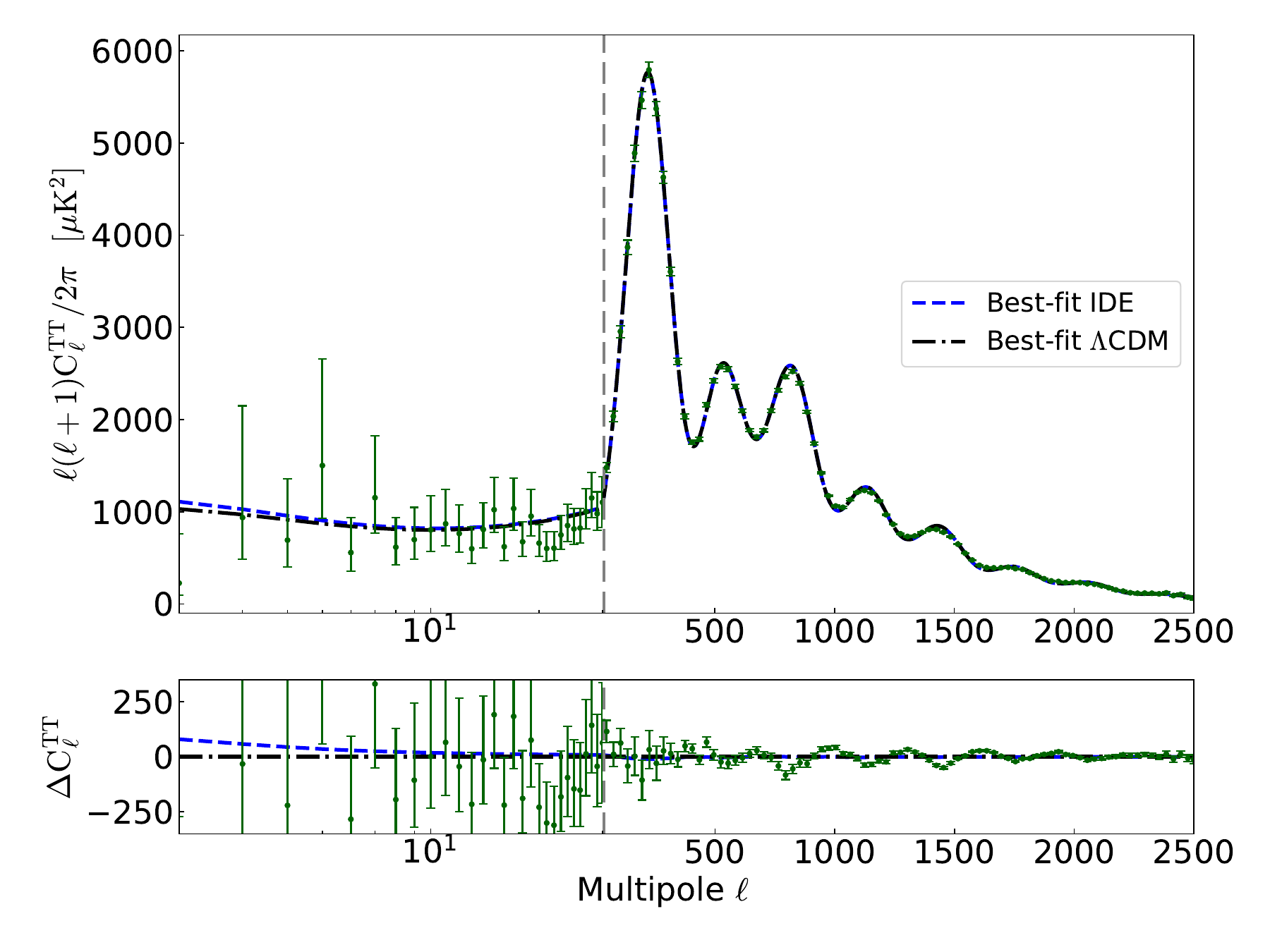} 
    \caption{\textbf{Left panel:} Statistical reconstruction of the rescaled expansion rate of the universe, $H(z)/(1+z)$, at 1$\sigma$ and 2$\sigma$ for the $\Lambda$CDM and IDE models, based on the joint analysis of CMB + DESI+ PP and  CMB + DESI+ PPS data respectively, compared to DESI measurements. \textbf{Right panel:} Comparison between the $\Lambda$CDM and IDE theoretical predictions for the temperature anisotropy power spectrum, with all cosmological parameters fixed to their respective best-fit values based on the CMB + DESI + PP joint analysis for $\Lambda$CDM and CMB + DESI + PPS for IDE. The error bars associated with the data points represent $\pm 1\sigma$ uncertainties. In the lower panel, we show the relative deviation between the IDE and $\Lambda$CDM theoretical predictions.}
    \label{fig:H(z)}
\end{figure*}

\textit{Assessing Consistency Across Multiple Supernova Samples}: We now proceed to evaluate the consistency of the model with respect to various SN Ia samples. Currently, three different SN Ia samples are publicly available in the literature, namely PP, Union3, and DESY5. The right side of Table~\ref{tab_results_I} shows the constraints for the IDE model in this work combining the three samples individually with CMB. Figure~\ref{fig:cmb_xi} in the right panel shows 1D and 2D contours at 68\% and 95\% CL in the plane $\xi_i-H_0$, illustrating the impact of these analyses. 

All analyses show a statistical trend favoring values of $\xi_i > 0$. Notably, the combination of CMB + PP data yields constraints on $\xi_i$ that are fully consistent with the null hypothesis, i.e., $\xi_i = 0$. In contrast, the CMB + Union3 dataset suggests a slight preference for $\xi_i > 0$, although this trend remains within the 1$\sigma$ level. The most significant result arises from the CMB + DESY5 combination, which indicates strong evidence for $\xi_i > 0$, reaching up to 3$\sigma$ confidence. This clearly demonstrates that different SN Ia samples can have a substantial impact on the observational constraints of $\xi_i$. It is interesting to note that the parametric space for the parameters $\xi_i$, $H_0$, and $S_8$, derived from the CMB + DESI and CMB + PPS combinations \footnote{By imposing a Gaussian prior on $M_B = -19.2435 \pm 0.0373$ \cite{Camarena_2021}, we confirm that the analyses using CMB + PPS and CMB + PP + $M_B$ are statistically equivalent. Consequently, all interpretations and results derived in this work under the CMB + PPS framework remain valid when incorporating CMB + PP + $M_B$.}, suggest possible evidence for $\xi_i < 0$. In contrast, the CMB + SN (PP, Union3, DESY5) datasets tend to favor $\xi_i > 0$. These joint analyses, which favor opposite signs for $\xi_i$, are in tension with each other according to Eq. (\ref{N-tension}), specifically, the CMB + PP analysis shows a tension of $2.3\sigma$ with CMB + DESI, CMB + Union3 exhibits a $2.6\sigma$ tension, and CMB + DESY5 reaches up to a $3.8\sigma$ tension. Due to these tensions, we chose not to combine the SN datasets (which favor $\xi_i > 0$) with DESI for a joint analysis to avoid potential bias in our analysis.

The tension observed between the uncalibrated PP data and DESI, which is resolved when using the calibrated PPS data, can be attributed to the new correlations introduced by the additional parameters in the IDE model, particularly $\xi$, which strongly affects $H_0$. The parameter $\xi$ is significantly correlated with $H_0$, and its sign can either increase or decrease $H_0$ values, as discussed earlier. The critical aspect lies in the role of the DESI sample. While the PP and other SN samples naturally predict lower values of $H_0$, driving $\xi$ to positive values, the DESI-BAO measurements—unlike previous BAO samples from the SDSS era—provide constraints in the $\xi$-$H_0$ plane that allow for relatively higher $H_0$ values (see ~\cite{Giare:2024smz} for details). This explains why the DESI data align well with the calibrated PPS dataset within the IDE framework. Since the PPS dataset is calibrated using SH0ES measurements, which predict higher $H_0$ values, the compatibility of the DESI sample with PPS is unsurprising. In the IDE context, the DESI sample also predicts slightly higher $H_0$ values.
Conversely, datasets such as PP, DESY5, and Union 3.0, which exhibit tension with PPS in $H_0$ even within the $\Lambda$CDM model, also show tension with DESI under the IDE framework due to their inherent preference for lower $H_0$ values.

While there is a general trend toward $\xi_i > 0$, influenced by the new correlations introduced by the model, this scenario is associated with lower values of $H_0$ and higher values of $S_8$ when compared to the standard $\Lambda$CDM predictions. In other words, although the datasets exhibit a statistical preference for $\xi_i > 0$, this scenario does not resolve, and may even worsen, the existing tensions in both $H_0$ and $S_8$. Despite these tensions, the scenario with $\xi_i > 0$ is still seen as providing a better fit compared to the standard $\Lambda$CDM model in certain respects. For joint analysis CMB + DESY5, we find $\Delta \chi^2_{\rm min} = -4.56$, $\Delta \text{AIC} = -2.56$ and, $\ln{\mathcal{B}_{ij}} = -1.26$.
It is interesting to note that, when including the SN Ia samples in a joint analysis, we find a predicted value of $z_{\rm eq, dark} \sim 0.32$, which is slightly higher than values derived from other analyses. This suggests that SN Ia data can have a significant impact on models with late-time transitions.

As noted in~\cite{Efstathiou:2024xcq}, a magnitude offset of approximately 0.04 mag between low and high redshifts was identified between the PantheonPlus and DESY5 samples. This discrepancy contributes to the observed deviations from $\Lambda$CDM cosmology when analyzing the DESY5 data. If this offset is not attributable to systematics in the DESY5 sample, our model indicates that a transition in the underlying dynamics may occur at late times, within redshift intervals corresponding to these magnitude differences. In such a scenario, models with $\xi_i > 0$ could provide an alternative phenomenological explanation for the observed discrepancy.


Figure~\ref{fig:H(z)} (left panel) presents a statistical reconstruction up to 2$\sigma$ for the expansion rate $H(z)$ in both the IDE model and $\Lambda$CDM, using the best-fit values and covariance matrix from the joint CMB + PPS + DESI analysis. Generally, in this case, the expansion rate $H(z)$ predicted by the IDE model is consistently higher than that predicted by $\Lambda$CDM. In the right panel, we present the theoretical effects on the temperature anisotropy power spectrum, with all cosmological parameters fixed at their respective best-fit values based on the CMB + DESI + PPS joint analysis. We observe differences between the two models only at large angular scales, where the best-fit IDE temperature anisotropy power spectrum exhibits a slightly greater amplitude compared to the baseline $\Lambda$CDM case. This difference arises primarily from the late integrated Sachs-Wolfe effect, which is attenuated due to the presence of dark coupling. Consequently, this alters the ratio between the densities of $\Omega_{\text{cdm}}(z)$ and $\Omega_x(z)$ at late times compared to the $\Lambda$CDM model, leading to the observed effects shown in the figure, since the other parameters, such as $A_s$ and $n_s$, are essentially identical between the IDE and $\Lambda$CDM models (see Table~\ref{tab_results_I}).

\section{Conclusions}
\label{conclusions}

Due to various observational tensions on some cosmological parameters assuming the $\Lambda$CDM model, cosmological models based on a non-gravitational interaction between DM and DE have been extensively explored in the recent literature as an alternative to the standard paradigm. Interacting scenarios involving energy-momentum transfer from DE to DM, and vice versa, have typically been investigated separately. In this work, we have proposed a more general framework where, in addition to the presence of a coupling in the dark sector throughout cosmic time, there is also an abrupt transition in the direction of energy-momentum transfer between the dark species at late times. In other words, the sign of the coupling parameter is allowed to change at low redshifts, which we name $z_{\rm eq, dark}$, marking a shift in the interaction dynamics.

This shift is motivated by the need to resolve persistent tensions in cosmological observations, particularly the $H_0$ and $S_8$ tensions, and may offer a more flexible mechanism for accommodating late-time observational data. However, such models are known to suffer from various theoretical instabilities, which can challenge their viability. To address these issues, we have developed a consistent numerical implementation that ensures the stability of the system while allowing for a change in the sign of the coupling parameter, as detailed in section~\ref{model}.

Our results show that the change in the sign of the coupling function in the dark sector can alter the correlations between cosmological parameters established in previous studies. Specifically, for models where $\xi_i < 0$ transitions to $\xi_i > 0$ at late times, this framework can predict higher values of $H_0$ and lower values of $S_8$ in a controlled manner. Scenarios with these key characteristics could be strong candidates for simultaneously resolving both tensions. Further, the new observational constraints are evaluated using the most recent and robust SN Ia samples available in the literature, which suggest that different SN Ia datasets can affect the constraints on the coupling parameter differently. The joint analysis CMB + PPS provides evidence for $\xi_i < 0$, whereas the CMB + DESY5 analysis suggests the opposite, favoring $\xi_i > 0$.

Another intriguing aspect to explore in future studies is the epoch of the transition. Since deriving a theoretical motivation from first principles is non-trivial, or even impossible to postulate at this moment, our robust joint analysis fixes the transition timing around the cosmic coincidence period, yielding $z_{\rm eq, dark} \sim 0.25$. However, we observe that $z_{\rm eq, dark}$ exhibits a negative correlation with $H_0$ and a positive correlation with $S_8$. Therefore, selecting different ad hoc transition times could potentially offer a complete and simultaneous solution to both tensions. 

Although the potential of interacting dark energy models to resolve the $S_8$ tension has been discussed previously in (see \cite{Murgia:2016ccp} for example), our model expands this well-established phenomenological framework. This extension has the potential to address both the $S_8$ and $H_0$ tensions simultaneously, while maintaining stable constraints on $\Omega_{\rm m}$ and $\sigma_8$ in comparison with measurements of large-scale structures, a success observed in only a few models proposed in recent literature \cite{Akarsu:2024eoo, Akarsu:2024qsi, Niedermann:2023ssr, Cruz:2023lmn, Naidoo:2022rda}. A limitation found in our results is that this pattern appears only when Cepheid distance calibration in the SH0ES samples is taken into account or when BAO-DESI data are included in the analysis, indicating the need for a broader range of observational tests to draw more definitive conclusions. When fitting other datasets and their combinations, our model does not show a statistically significant preference for higher values of $H_0$, but in the presence of some recent SN Ia samples, it can lead to $\xi_i > 0$.
\\

\textbf{Data Availability}: The datasets and products underlying this research will be available upon reasonable request to the corresponding author after the publication of this article.
\\

\begin{acknowledgments}
\noindent The authors express their gratitude to the referee for the valuable comments and suggestions, which have contributed to enhancing the significance of the results presented in this work. 
The authors thank William Giarè for thorough discussions during the development of the early stages of this work. M.A.S and E.S. received support from the CAPES scholarship. R.C.N. thanks the financial support from the Conselho Nacional de Desenvolvimento Científico e Tecnologico (CNPq, National Council for Scientific and Technological Development) under the project No. 304306/2022-3, and the Fundação de Amparo à Pesquisa do Estado do RS (FAPERGS, Research Support Foundation of the State of RS) for partial financial support under the project No. 23/2551-0000848-3.  S.K. gratefully acknowledges the support of Startup Research Grant from Plaksha University  (File No. OOR/PU-SRG/2023-24/08).
E.D.V. is supported by a Royal Society Dorothy Hodgkin Research Fellowship. This article is based upon work from the COST Action CA21136 - ``Addressing observational tensions in cosmology with systematics and fundamental physics (CosmoVerse)'', supported by COST - ``European Cooperation in Science and Technology''.

\end{acknowledgments}

\bibliographystyle{apsrev4-1}
\bibliography{main}

\end{document}